\documentclass{aastex631}
\usepackage{xcolor}
\usepackage{subfigure}
\usepackage{amsmath}
\usepackage{color, soul}
\usepackage{bm,amsmath}
\usepackage{dcolumn}
\usepackage{bm}
\usepackage{epsf}
\usepackage{verbatim}
\usepackage{amssymb}
\usepackage{threeparttable}
\usepackage{multirow}
\usepackage{verbatim} 
\usepackage[normalem]{ulem} 
\usepackage{mathtools}
\usepackage{amsfonts}
\usepackage{relsize}
\setstcolor{red}

\usepackage{booktabs} 

\newcommand{\approptoinn}[2]{\mathrel{\vcenter{
  \offinterlineskip\halign{\hfil$##$\cr
    #1\propto\cr\noalign{\kern2pt}#1\sim\cr\noalign{\kern-2pt}}}}}

\shorttitle{Wind from a massive white dwarf merger product -- II.}
\shortauthors{Zhong et al.}
\graphicspath{{./}{figures/}}

\begin{document}

\title{The optically thick rotating magnetic wind from a massive white dwarf merger product -- II. 
axisymmetric magnetohydrodynamic simulations}

\author[0000-0003-0805-8234]{Yici Zhong}
\email{yici.zhong@phys.s.u-tokyo.ac.jp}
\affiliation{Department of Physics, Graduate School of Science, University of Tokyo, Bunkyo-ku, Tokyo 113-0033, Japan}

\author[0000-0003-4299-8799]{Kazumi Kashiyama}
\affiliation{Research Center for the Early Universe, Graduate School of Science, University of Tokyo, Bunkyo-ku, Tokyo 113-0033, Japan}
\affiliation{Kavli Institute for the Physics and Mathematics of the Universe (Kavli IPMU,WPI), The University of Tokyo, Chiba 277-8582, Japan}

\author[0000-0003-3882-3945]{Shinsuke Takasao}
\affiliation{Department of Earth and Space Science, Graduate School of Science, Osaka University, Toyonaka, Osaka 560-0043, Japan}

\author[0000-0002-4060-5931]{Toshikazu Shigeyama}
\affiliation{Research Center for the Early Universe (RESCEU), School of Science, The University of Tokyo, 7-3-1 Hongo, Bunkyo-ku, Tokyo 113-0033, Japan}
\affiliation{Department of Astronomy, School of Science, The University of Tokyo, 7-3-1 Hongo, Bunkyo-ku, Tokyo 113-0033, Japan}

\author[0000-0002-9072-4744]{Kotaro Fujisawa}
\affiliation{Department of Liberal Arts, Tokyo University of Technology, Ota-ku, Tokyo 144-0051, Japan}
\affiliation{Research Center for the Early Universe (RESCEU), School of Science, The University of Tokyo, 7-3-1 Hongo, Bunkyo-ku, Tokyo 113-0033, Japan}

\date{\today}


\begin{abstract}
We numerically construct a series of axisymmetric rotating magnetic wind solutions, aiming at exploring the observation properties of massive white dwarf (WD) merger remnants with a strong magnetic field, a fast spin, and an intense mass loss, as inferred for WD J005311. 
We investigate the magnetospheric structure and the resultant spin-down torque exerted to the merger remnant with respect to the surface magnetic flux $\Phi_*$, spin angular frequency $\Omega_*$ and the mass loss rate $\dot M$.
We confirm that the wind properties for $\sigma \equiv \Phi^2_* \Omega_*^2/\dot M v_\mathrm{esc}^3 \gtrsim 1$ significantly deviate from those of the spherical Parker wind, where $v_\mathrm{esc}$ is the escape velocity at stellar surface. 
For such a rotating magnetic wind sequence, we find:
(i) quasi-periodic mass eruption triggered by magnetic reconnection along with the equatorial plane
(ii) a scaling relation for the spin-down torque $T \approx (1/2) \times \dot{M} \Omega_* R^2_* \sigma^{1/4}$. 
We apply our results to discuss the spin-down evolution and wind anisotropy of massive WD merger remnants, the latter of which could be probed by a successive observation of WD J005311 using {\it Chandra}.
\end{abstract}

\keywords{white dwarfs --- stars: winds, outflows --- stars: rotation}


\section{Introduction} \label{sec:intro}
Consequences of a merger of massive white dwarfs (WDs) are of great astrophysical importance.
It may explode as a Type Ia supernova in particular when the binary constitutes of carbon-oxygen WDs with a total mass exceeding the Chandrasekhar limit~\citep{1984ApJ...277..355W, 1984ApJS...54..335I}. 
Instead, if a super-Chandrasekhar oxygen-neon core is synthesized after the merger, it may collapse into a neutron star~(NS)~\citep[][]{1985ApJ...297..531N, 2004ApJ...615..444S}. Such a merger induced collapse has gotten attention as a scenario for the formation of peculiar type of neutron stars, e.g., sources of fast radio bursts~\citep[e.g.,][]{2017ApJ...839L...3K,2021ApJ...917L..11K,2022Natur.602..585K,2022MNRAS.510.1867L}. 

If not explode nor collapse, the merger product will be a rapidly rotating and strongly magnetized WD~\citep[e.g.,][]{2008MNRAS.387..897T, 2015MNRAS.447.1713B}. They would constitue a good fraction, say $\sim 20$ \%, of the Galactic massive WDs with a mass of $M_* \gtrsim 1\,M_\odot$~\citep[e.g.,][]{2012ApJ...749...25G,2020ApJ...891..160C,2021ApJ...906...53S}.
Thanks to rather complete photometric searches and spectroscopic followups, increasing amount of merged WD candidates have been identified, e.g., ZTF J190132.9+145808.7 with $M_* = (1.327\mbox{-}1.365)M_\odot$, $P = 6.97\,\mathrm{min}$ and $B_* = (6\mbox{-}9)\times 10^{8}\,\mathrm{G}$~\citep[][]{2021Natur.595...39C} and SDSS J221141.80+113604.5 with $M_* = 1.268\,M_\odot$, $P = 76\,\mathrm{sec}$ and $B_* = 1.5\times 10^{7}\,\mathrm{G}$~\citep[][]{2021ApJ...923L...6K}, where $P$ and $B_*$ denote the spin period and the strength of the surface magnetic field at the pole. Their post-merger ages have been estimated as $\sim 10\,\mathrm{Myr}$ and $\sim 100\,\mathrm{Myr}$, respctively, from their positions on the cooling track. 

Recently, a candidate for a significantly younger merger product, WD J005311, was fortuitously discovered within an infrared nebula~\citep[][]{2019Natur.569..684G}. The most remarkable characteristic of this WD is unveiled through optical spectroscopy, revealing an optically-thick wind emanating from it. This wind is enriched with carbon burning ashes and exhibits a remarkable velocity of $v_{\infty}=16,000 \pm 1,000 \mathrm{~km} \mathrm{~s}^{-1}$, accompanied by a mass loss rate of $\dot{M}=(3.5 \pm 0.6) \times 10^{-6} M_{\odot} \, \mathrm{yr}^{-1}$. While the direct measurement of the central WD's physical properties remains elusive, the presence of such a fast and intense wind strongly suggests that it is a rapidly rotating and strongly magnetized WD, potentially possessing a super- or near-Chandrasekhar mass~\citep{2019Natur.569..684G,2019ApJ...887...39K}.

The mass and composition loaded on the WD J005311 wind is likely from the near-surface carbon burning. The launch of such a wind can be triggered by the Kelvin-Helmholtz contraction of the oxygen neon core of the merged WD, that can happen $\sim 1,000\mbox{-}10,000\,\mathrm{yr}$ after the merger~\citep{2016MNRAS.463.3461S,2023MNRAS.tmp.1861Y,2023ApJ...944L..54W}. The timing can be consistent with the post-merger age of the system estimated based on both the expansion velocity of the surrounding nebula and the ancient records on a historical Galactic SN, SN1181, which happened in the direction of WD J005311 $\sim 850\,\mathrm{yr}$ ago and is likely associated with the merger of the progenitor binary~\citep{2021ApJ...918L..33R,Lykou2022,2023arXiv230414669K}.

On the other hand, the expansion velocity of the wind observed in WD J005311 significantly surpasses the escape velocity of a WD with a typical mass. This suggests that the wind is either thermally driven, originating from a super- or near-Chandrasekhar mass WD, or magnetically driven due to the rapid rotation and strong magnetic field of the WD. In the former case, the wind velocity will be~\citep{1965SSRv....4..666P}:
\begin{equation} \label{eq:michel}
v_\mathrm{T} \approx \sqrt{\frac{2GM_*}{R_*}} \sim 20,000\,\mathrm{km\,s^{-1}}\,\left(\frac{M_*}{1.4\,M_\odot}\right)^{1/2}\left(\frac{R_*}{1,000\,\mathrm{km}}\right)^{-1/2},
\end{equation}
while in the latter case, the maximum wind velocity along the equatorial plane is~\citep{1967ApJ...148..217W,1969ApJ...158..727M}:
\begin{equation} \label{eq:WD}
v_\mathrm{M, max} \approx \left(\frac{B_*^2 R_*^4\Omega_*^2}{\dot M}\right)^{1/3} \sim 13,000\,\mathrm{km\,s^{-1}}\left(\frac{B_*}{2\times 10^7\,\mathrm{G}}\right)^{2/3}\left(\frac{R_*}{4,000\,\mathrm{km}}\right)^{4/3}\left(\frac{\Omega_*}{0.2\,\mathrm{s^{-1}}}\right)^{2/3}\left(\frac{\dot M}{3\times 10^{-6}\,M_\odot\,\mathrm{yr^{-1}}}\right)^{-1/3}.
\end{equation}
The wind is so fast that it catches up and clashes into the surrounding supernova ejecta, forming a wind termination shock, which is observed as an inner X-ray nebula~\citep{Oskinova_et_al_20,2023arXiv230414669K}. The X-ray nebula is still in its infancy; given the observed angular size, it is only a few tens of years old~\citep[][]{2023arXiv230414669K}. Subsequent observations may reveal the time variability and anisotropy of the wind, which is generally expected for a rotating magnetic wind but has not been explored in this context. These properties of the wind can also be linked to the mass-loss and spin-down rates of the central WD, which are important in determining the fate of the central WD: whether it eventually collapses into a neutron star, and if so, how rapidly rotating and strongly magnetized the neutron star would be.

Here we model a system like WD J005311 by numerically constructing a 2D axisymmetric wind solution driven by rotating dipole, with implementing a wind launching region that mimics the near-surface carbon burning region. We investigate the wind structure together with its time evolution (i.e., how the mass, energy and angular momentum loss rate from the system evolves with time), and the scaling of the spin-down torque with respect to system parameters such as surface magnetic field, rotation frequency and mass loss rate. This paper is organized as follows. We introduce our setup in Sec. \ref{sec:setup}, including numerical details. In Sec. \ref{sec:result}, we show our results on wind structure, time evolution and scaling of spin-down torque. Finally, we discuss several implications and applications on observational results in Sec. \ref{sec:dis}.

\section{Setup} \label{sec:setup}

We conduct a series of numerical simulations of a rotating magnetic wind from a massive WD merger product with a stable nuclear burning occurring at the near surface region. 
We first describe the general numerical setup including the governing equations, the Riemann solver, the mesh decomposition, and the boundary conditions in Sec. \ref{numerical_detail}. We then describe the source term that represents the injection of mass and internal energy at the near surface nuclear burning region in Sec. \ref{BC}. Finally, we elaborate on setups related to magnetic fields.

\subsection{Magnetohydrodynamic (MHD) equations} 
\label{numerical_detail}
We numerically integrate ideal MHD equations with central gravity;
\begin{equation} \label{eq:masscons}
\frac{\partial \rho}{\partial t}+\nabla \cdot(\rho \mathbf{v}) =S_\rho,
\end{equation}
\begin{equation}
\frac{\partial (\rho \mathbf{v})}{\partial t}+\nabla \cdot \mathbf{T}=-\rho \nabla \phi,
\end{equation}
\begin{equation}
\frac{\partial \varepsilon_\mathrm{tot}}{\partial t}+\nabla \cdot \mathbf{s} =S_\mathrm{e} - \rho (\nabla \phi \cdot \mathbf{v}),
\end{equation}
\begin{equation} \label{eq:Btheta}
\frac{\partial \mathbf{B}}{\partial t} - \nabla \times (\mathbf{v} \times \mathbf{B})=0,
\end{equation}
in the two dimensional spherical coordinate using \texttt{Athena++} \footnote{\url{https://github.com/PrincetonUniversity/athena}}~\citep[][]{Stone2020}. Here $S_\rho$ and $S_\mathrm{e}$ are source terms that we use to mimic the matter and energy injection into the computational domain, which will be described in detail in Sec. \ref{BC}; the velocity vector $\mathbf{v}$, magnetic field $\mathbf{B}$, stress tensor $\mathbf{T}$, total energy density $\varepsilon_\mathrm{tot}$, energy flux $\mathbf{s}$ are given as
\begin{equation}
    \mathbf{v} =\left(v_r,v_\theta, v_\varphi\right)\left(\begin{array}{c}
\hat{\boldsymbol{r}} \\
\hat{\boldsymbol{\theta}} \\
\hat{\boldsymbol{\varphi}},
\end{array}\right)
\end{equation}

\begin{equation}
    \mathbf{B} =\left(B_r,B_\theta, B_\varphi\right)\left(\begin{array}{c}
\hat{\boldsymbol{r}} \\
\hat{\boldsymbol{\theta}} \\
\hat{\boldsymbol{\varphi}}
\end{array}\right)
\end{equation}

\begin{equation} \label{tenserT}
\mathbf{T} =\rho \mathbf{v}  \mathbf{v}+\left(p+\frac{|\mathbf{B}|^2}{8 \pi}\right) {\mathbf{I}}-\frac{\mathbf{B}  \mathbf{B}}{4 \pi},
\end{equation} 

\begin{equation} \label{ene_den}
\varepsilon_\mathrm{tot} =\frac{\rho |\mathbf{v}|^2}{2}+\frac{|\mathbf{B}|^2}{8 \pi}+\frac{p}{\gamma-1},
\end{equation}

\begin{equation}
\mathbf{s} = \left(\frac{1}{2} \rho |\mathbf{v}|^2+ \frac{\gamma}{\gamma-1} p + \frac{|\mathbf{B}|^2}{4 \pi} \right) \mathbf{v}-\frac{\mathbf{B} (\mathbf{v} \cdot \mathbf{B})}{4 \pi},
\end{equation}
where $\rho$ is the density, $p$ is the pressure, ${\mathbf{I}}$ is the identity dyadic tensor,
and $\phi = -GM_*/r$ is the gravitational potential, where $G$ is the gravitational constant, $M_*$ is the mass of the central WD. 
To close Eqs.(\ref{eq:masscons})-(\ref{eq:Btheta}), we use the adiabatic equation of state with an index of $\gamma = 4/3$. 
The above ideal MHD equations are scale-free; we use a unit of $G= M_* = R_* =1$ for the numerical calculations, where $R_*$ is the radius of the WD. 
When estimating quantities in a physical unit, we transform to the cgs unit with setting $M_* = 1 \, M_\odot$ and radius $R_* = 0.009~\,R_\odot$. We note that this is consistent with the mass-radius relation of degenerate oxygen neon cores with an angular frequency of $\Omega_* \lesssim 0.5\,\mathrm{s^{-1}}$~\citep{2019ApJ...887...39K}.  

We use the HLLD approximate Riemann solver for the MHD equations~\citep[][]{2005JCoPh.208..315M} with the second-order piecewise linear reconstruction method (PLM). The time integration is carried out by the second-order Runge-Kutta method with Courant-Friedrich-Lewy number of 0.1. The computational domain is resolved with the mesh number of 128 for [$0.9,30$] $R_*$ in the radial direction and 128 for [$0, \pi$] in the polar direction. We employ a non-uniform mesh in the radial direction, where the radial grid size is proportional to the radius. The fiducial value of the grid size ratio $\Delta r(i+1)/\Delta r(i) $ is 1.02 so that the smallest cell size is 0.05, where $i$ stands for the grid index.

At the outer boundary of the computational domain, we impose the zero-gradient boundary condition for the radial direction and connect the domain across the axes for the polar direction.
On the other hand, we impose the zero gradient boundary condition for the inner boundary ($r = r_\mathrm{in} = 0.9 R_*$), and set the velocity to be compatible with the rigid rotation of the central WD;
    \begin{equation}
        v_{r,\mathrm{in}}=v_{\theta,\mathrm{in}}=0, \ \ v_{\varphi,\mathrm{in}} = \Omega_* r_\mathrm{in} \sin \theta.
    \end{equation}
In this paper, we consider the cases with $\Omega_* = [0.05,0.07,0.12,0.16,0.23,0.35,0.46]\,\mathrm{s^{-1}}$, which correpsonds to $\sim 5\mbox{-}50$ \% of the mass shedding limit. In terms of the inner ghost cell's density and pressure, we carefully prescribe their values to achieve a specific thermally-driven wind mass loss rate (see Sec. \ref{BC}). Initially, we distribute a cold and homogeneous gas throughout the entire computational domain and inject the thermally-driven wind from the designated launching region. As the thermally-driven outflow reaches the outer boundary, we initiate an aligned dipole field at the inner boundary, facilitating the transformation of the wind into a rotating magnetic wind (see Sec. \ref{wlr}).

\subsection{Wind launching region} \label{BC}

{We initialize our simulation with a cold, homogeneous, isotropic and non-magnetized atmosphere, }{and set up a ``wind launching region''~\footnote{Note that this is originally called damping layer in the context of accreting stellar system~\citep[see][]{2019ApJ...878L..10T}} with a width of $\mathcal{D}$ near the WD surface, where the mass is injected to the computational domain to
mimic the mass loading due to the carbon burning around the surface of massive WD merger product.
To do that, we implement an isotropic relaxation function for both matter and energy source terms to update density and pressure in wind launching region}:
\begin{equation}\label{eq:source_term_rho}
    S_\rho = \frac{\rho_{*}-\rho}{\tau}, \ \ \ (r_\mathrm{in} \leq r  \leq \mathcal{D}), 
\end{equation}
\begin{equation}\label{eq:source_term_e}
    S_\mathrm{e} = \frac{p_{*}-p}{\tau}, \ \ \ (r_\mathrm{in} \leq r  \leq \mathcal{D}), 
\end{equation}
where $\rho_*$ and $p_*$ correspond to the density and pressure at the outer edge of the wind launching region. The actual value of the relaxation timescale $\tau$ is chosen to satisfy the condition,
\begin{equation} \label{eq:tau}
    \tau \lesssim \frac{\mathcal{D}}{\max{(v_\mathrm{s, *}, v_\mathrm{A,*})}},
\end{equation}
where $v_\mathrm{s,*} \equiv \sqrt{\gamma p_* / \rho_*}$ is the adiabatic sound velocity and $v_\mathrm{A,*} \equiv \sqrt{B_*^2 / 4 \pi \rho_*}$ is the Alfv$\acute{\text{e}}$n velocity with $B_*$ being the surface magnetic field strength at the equator (see Sec. \ref{wlr}). 
This condition is needed to stably inject mass to the computational domain by suppressing fluctuations associated with hydrodynamic and/or MHD waves in the wind launching region. 
The above source terms can self-consistently produce a thermal pressure-driven wind with $\rho \propto r^{-2}$, $v_{r} \approx v_\mathrm{esc}$ and a stable mass loss in the steady state, where $v_\mathrm{esc}$ is the surface escape velocity. 
In this paper, we set the width of the wind launching region as ${\cal D} = 0.6 ,R_*$ as our fiducial value and check the convergence of our results with respect to the value of ${\cal D}$.
We use a fixed value of $\tau$, with which Eq. (\ref{eq:tau}) is satisfied for the most strongly magnetized case. Then we set $\rho_*$ and $p_*$ so that the mass loss rate by the thermal pressure-driven wind becomes $\dot{M}$ = $10^{-6}$ $M_\odot$ $\mathrm{yr}^{-1}$.
    
\subsection{Rotating magnetic wind} \label{wlr}

{After the thermal pressure-driven wind settles down, }we turn on a dipole magnetic field that is embedded on the rotating stellar surface, with magnetic moment $\boldsymbol{\mu} \equiv B_* R^3_* \hat{\boldsymbol{z}}$ aligned with the rotation axis. We use the following vector potential
\begin{equation}\label{eq:vecA}
\begin{split}
    \mathbf{A}(\mathbf{r}) & = \left(\frac{\boldsymbol{\mu} \times \mathbf{r}}{r^{3}}\right)_d \\
    & = \left(0, 0, \frac{B_* R^3_* \sin \theta }{r^2}\right) \left(\begin{array}{c}
\hat{\boldsymbol{r}} \\
\hat{\boldsymbol{\theta}} \\
\hat{\boldsymbol{\varphi}}
\end{array}\right)
\end{split}
\end{equation}
to ensure that the divergence of magnetic field vanishes. We consider the cases with $B_* = $ [$1.5\times 10^6, 2.3\times 10^6, 2.6\times 10^6, 3.0\times 10^6$] G, for which $\beta_* \equiv v_\mathrm{s,*}/v_\mathrm{A, *} = 10^{-(2-3)} \ll 1$ so that the magnetic pressure dominates in the near surface region. 
In order to numerically solve the MHD equations in such a low plasma beta gas, we implement the dual energy formalism (see Appendix \ref{appen:dual_e}).

The launched gas will then corotate with the rotating magnetic field, and the magnetic torque, which depends on the resultant magnetospheric structure and the polar angle, can also contribute to the wind acceleration in addition to the thermal pressure gradient. As a result, we expect a rotating magnetic wind to start blowing in an angle dependent manner, and relax to a quasi-steady state when it reaches to the outer boundary. We simulate the rotating magnetic wind for a few 10 $\times$ the spin period after turning on the magnetic field.

\section{Result} \label{sec:result}

 Table. \ref{table:3} shows a summary of our simulation. A model $\tt{B x \Omega y}$ corresponds to the case with $B_* = x$ and $\Omega_* = y$ in the cgs unit. When the rotating magnetic wind becomes quasi-steady, it can be characterized by the mass loss rate 
\begin{equation} \label{eq:mdot}
    \dot{M}={2 \pi} \int^{\pi}_0 \rho v_r r^2 \sin \theta d\theta,
    \end{equation}
wind luminosity
\begin{equation} \label{eq:windL}
    L={2 \pi} \int^{\pi}_0 \rho v_r r^2 \left[\frac{1}{2} v^2 + \frac{\gamma P}{\rho(\gamma-1)} - \frac{\Omega_* r \sin \theta B_r B_\phi}{4 \pi \rho v_r} \right]  \sin \theta d\theta,
    \end{equation}
and spindown torque
\begin{equation} \label{eq:torque}
    T = 2 \pi \int^{\pi}_0 \rho v_r r^2 \left(r v_\phi - \frac{r B_r B_\phi}{4 \pi \rho v_r}\right) \sin \theta d\theta
    \end{equation}
estimated  at the outer boundary. As we show later, the strength of rotating magnetic winds can be characterized by a dimensionless parameter  
\begin{equation} \label{sigma}
    \sigma \equiv \frac{ \Phi_*^2\Omega_*^2 }{\dot{M}v_\mathrm{esc}^3},
\end{equation}
where $\Phi_* \equiv {2 \pi} \int^{\pi/2}_0 B_r r^2 \sin \theta d \theta |_{r = R_*}$ is the half hemisphere magnetic flux and $v_\mathrm{esc} = \sqrt{{2GM_*}/{R_*}}$ is the escape velocity at the WD surface~\footnote{In relativistic MHD regime, speed of light $c$ is conventionally used as the characteristic speed of the system~\citep[e.g., see the definition of $\sigma_0$ in][]{2006MNRAS.368.1717B}.}.
With using $\sigma$, the Michel velocity (Eq.~\ref{eq:michel}) can be described as $v_{\mathrm{M, \max}} \approx \sigma^{1/3} v_\mathrm{esc}$. 
Our simulations cover the range of $1 \lesssim \sigma \lesssim 500$. 

Hereafter we take $\tt B1.5e6\Omega0.23$ with $\sigma=34.3$
as the fiducial model, and first show the multi-dimensional structure of the rotating magnetic wind in Sec. \ref{MSwindstru}. We then investigate the time variability of the system primarily focusing on the impacts of quasi-periodic eruption along with the equatorial plane in Sec. \ref{EML}. Finally, we show how the time-averaged spin-down torque scales with system parameters in Sec. \ref{spindowntorque}.

\begin{table}[ht]
\centering
{\bf{Table 1.}} Summary of our simulations for rotating magnetic winds from white dwarfs
\begin{tabular*}{0.73\linewidth}{c|cc|cccc }
\hline \hline 
\multicolumn{1}{c|}{} & \multicolumn{2}{c|}{input parameters} & \multicolumn{4}{c}{calculated quantities$^{\dagger}$}\\
\cline{2-7}
\text {Model} & \text { $B_*$ [G]$^{a}$ } & $\Omega_*$ [$\mathrm{s}^{-1}$]$^{b}$ & $\dot{M}$ [$M_\odot$ $\mathrm{yr}^{-1}$]$^{c}$ & $L$ [$\mathrm{erg}$ $\mathrm{s}^{-1}$] $^{d}$& $T$ [dyn cm] $^{e}$& $\sigma$ $^{f}$\\
\hline 

 $\tt B3.0e6\Omega0.46$ & $3.0 \times 10^{6}$
 & $0.46$ & $1.77 \times 10^{-6}$ & $7.70 \times 10^{37}$ & $1.80 \times 10^{38}$ & $382$ \\ 
 
 $\tt {B2.6e6\Omega0.46}$ & $2.6 \times 10^{6}$
 & $0.46$ & $1.62 \times 10^{-6}$ & $6.83 \times 10^{37}$ & $1.70 \times 10^{38}$ & $317$ \\ 
 
 $\tt B2.3e6\Omega0.46$ & $2.3 \times 10^{6}$
 & $0.46$ & $1.52 \times 10^{-6}$ & $6.06 \times 10^{37}$ & $1.57 \times 10^{38}$ & $258$ \\
 
 $\tt B1.5e6\Omega0.46$ & $1.5 \times 10^{6}$
 & $0.46$ & $1.50 \times 10^{-6}$ & $5.44 \times 10^{37}$ & $1.33 \times 10^{38}$ & $122$ \\
 \hline 
 
 $\tt B3.0e6\Omega0.35$ & $3.0 \times 10^{6}$
 & $0.35$ & $1.65 \times 10^{-6}$ & $6.00 \times 10^{37}$ & $1.21 \times 10^{38}$ & $235$ \\
 
 $\tt B2.6e6\Omega0.35$ & {$2.6 \times 10^{6}$}
 & {$0.35$} & {$1.49 \times 10^{-6}$} & {$5.10 \times 10^{37}$} & {$1.06 \times 10^{38}$}  & {$199$} \\
 
 $\tt B2.3e6\Omega0.35$ & {$2.3 \times 10^{6}$}
 & {$0.35$} & {$1.40 \times 10^{-6}$} & {$4.59 \times 10^{37}$} & {$9.72 \times 10^{37}$} & {$161$} \\
 
 $\tt B1.5e6\Omega0.35$ & $1.5 \times 10^{6}$
 & $0.35$ & $1.39 \times 10^{-6}$ & $4.13 \times 10^{37}$ & $7.91 \times 10^{37}$ & $75.8$

\\\hline $\tt B3.0e6\Omega0.23$ & $3.0 \times 10^{6}$
 & $0.23$ & $1.41 \times 10^{-6}$ & $4.18 \times 10^{37}$ & $5.23 \times 10^{37}$ & $111$ \\
 
 $\tt B2.6e6\Omega0.23$ & $2.6 \times 10^{6}$
 & $0.23$ & $1.27 \times 10^{-6}$ & $3.58 \times 10^{37}$ & $4.78 \times 10^{37}$ & $94.0$ \\
 
 $\tt B2.3e6\Omega0.23$ & $2.3 \times 10^{6}$
 & $0.23$ & $1.20 \times 10^{-6}$ & $3.23 \times 10^{37}$ & $4.31 \times 10^{37}$ & $76.2$ \\
 
 $\tt B1.5e6\Omega0.23$ & $1.5 \times 10^{6}$
 & $0.23$ & $1.24 \times 10^{-6}$ & $3.13 \times 10^{37}$ & $3.54 \times 10^{37}$ & $34.3$ \\
 \hline 
 
 $\tt B3.0e6\Omega0.16$ & $3.0 \times 10^{6}$
 & $0.16$ & $1.29 \times 10^{-6}$ & $3.40 \times 10^{37}$ & $2.80 \times 10^{37}$ & $52.5$ \\
 
 $\tt B2.6e6\Omega0.16$ & $2.6 \times 10^{6}$
 & $0.16$ & $1.13 \times 10^{-6}$ & $2.85 \times 10^{37}$ & $2.52 \times 10^{37}$ & $45.4$ \\
 
 $\tt B2.3e6\Omega0.16$ & $2.3 \times 10^{6}$
 & $0.16$ & $1.07 \times 10^{-6}$ & $2.59 \times 10^{37}$ & $2.31 \times 10^{37}$ & $36.4$ \\
 
 $\tt B1.5e6\Omega0.16$ & $1.5 \times 10^{6}$
 & $0.16$ & $1.11 \times 10^{-6}$ & $2.61 \times 10^{37}$ & $1.70 \times 10^{37}$ & $16.3$ \\
 \hline 
 
 $\tt B3.0e6\Omega0.12$ & $3.0 \times 10^{6}$
 & $0.12$ & $1.14 \times 10^{-6}$ & $2.84 \times 10^{37}$ & $1.25 \times 10^{37}$ & $20.7$ \\
 
 $\tt B2.6e6\Omega0.12$ & $2.6 \times 10^{6}$
 & $0.12$ & $1.00 \times 10^{-6}$ & $2.36 \times 10^{37}$ & $1.09 \times 10^{37}$ & $17.8$ \\
 
 $\tt B2.3e6\Omega0.12$ & $2.3 \times 10^{6}$
 & $0.12$ & $1.00 \times 10^{-6}$ & $2.14 \times 10^{37}$ & $9.86 \times 10^{36}$ & $14.5$ \\
 
 $\tt B1.5e6\Omega0.12$ & $1.5 \times 10^{6}$
 & $0.12$ & $1.00 \times 10^{-6}$ & $2.28 \times 10^{37}$ & $7.74 \times 10^{36}$ & $6.19$ \\
\hline 

 $\tt B3.0e6\Omega0.07$ & $3.0 \times 10^{6}$
 & $0.07$ & $1.08 \times 10^{-6}$ & $2.67 \times 10^{37}$ & $7.26 \times 10^{36}$ & $11.1$ \\
 
 $\tt B2.6e6\Omega0.07$ & $2.6 \times 10^{6}$
 & $0.07$ & $1.00 \times 10^{-6}$ & $2.19 \times 10^{37}$ & $6.39 \times 10^{36}$ & $9.62$ \\
 
 $\tt B2.3e6\Omega0.07$ & $1.5 \times 10^{6}$
 & $0.07$ & $1.00 \times 10^{-6}$ & $2.00 \times 10^{37}$ & $5.83 \times 10^{36}$ & $7.71$ \\
 
 $\tt B1.5e6\Omega0.07$ & $1.5 \times 10^{6}$
 & $0.07$ & $1.00 \times 10^{-6}$ & $2.21 \times 10^{37}$ & $4.92 \times 10^{36}$ & $3.21$ \\
 \hline 
 
 $\tt B3.0e6\Omega0.05$ & $3.0 \times 10^{6}$
 & $0.05$ & $1.00 \times 10^{-6}$ & $2.56 \times 10^{37}$ & $4.42 \times 10^{36}$ & $5.59$ \\
 
 $\tt B2.6e6\Omega0.05$ & $2.6 \times 10^{6}$
 & $0.05$ & $1.00 \times 10^{-6}$ & $2.10 \times 10^{37}$ & $3.99 \times 10^{36}$ & $4.86$ \\
 
 $\tt B2.3e6\Omega0.05$ & $1.5 \times 10^{6}$
 & $0.05$ & $1.00 \times 10^{-6}$ & $1.91 \times 10^{37}$ & $3.76 \times 10^{36}$ & $3.89$ \\
 
 $\tt B1.5e6\Omega0.05$ & $1.5 \times 10^{6}$
 & $0.05$ & $1.00 \times 10^{-6}$ & $2.16 \times 10^{37}$ & $3.25 \times 10^{36}$ & $1.59$ \\
\hline
\end{tabular*}

\begin{tablenotes}[para,flushleft,online,normal] 
    \item[a] Surface magnetic field strength (Eq. \ref{eq:vecA});
    \item[b] Spin angular frequency of the white dwarf;
    \item[c] Mass loss rate (Eq. \ref{eq:mdot});
    \item[d] Spin-down luminosity (Eq. \ref{eq:windL});
    \item[e] Spin-down torque (Eq. \ref{eq:torque});
    \item[f] Magnetization parameter (Eq. \ref{sigma});
    \item[$\dagger$] Time averaged values. 
\end{tablenotes}

\label{table:3}
\end{table}

\subsection{Anisotropic wind structure} \label{MSwindstru}

\begin{figure}
    \centering
    \includegraphics[width=0.8\textwidth]{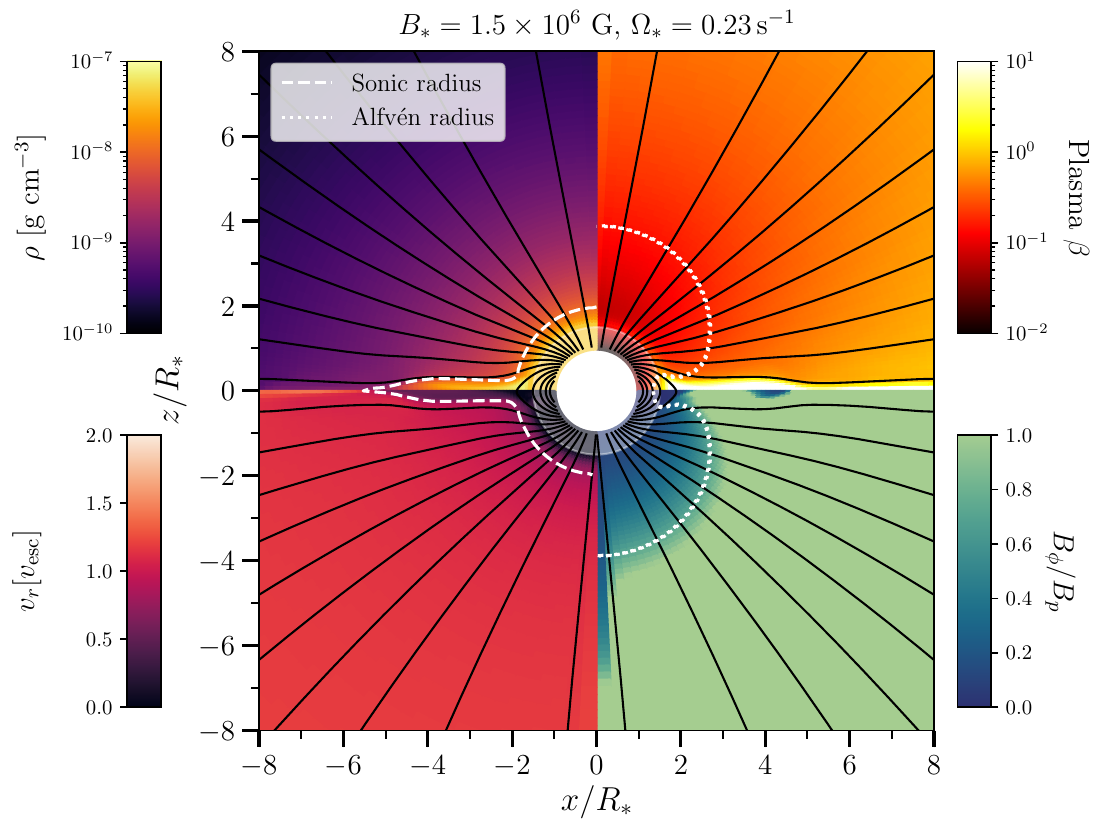}
    \caption{Snapshot of the rotating magnetic wind of $\tt B1.5e6\Omega0.23$ showing the gas density (left top), the plasma beta (right top), the radial velocity normalized by the escape velocity from the WD, and the ratio of the strength of toroidal field over poloidal field (right bottom). The solid, dashed, and dotted lines indicate the poloidal magnetic field lines, the position where the radial velocity of the wind exceeds the adiabatic sound velocity and the Alfv$\acute{\text{e}}$n velocity, respectively. The pale shaded region around the WD surface corresponds to the wind launching region.}
    \label{fig:wind_profile}
\end{figure}

Fig.~\ref{fig:wind_profile} shows a snapshot of our fiducial model ($\tt B1.5e6\Omega0.23$) after the wind structure reaches a quasi-steady state. 
As explained in Sec. \ref{sec:setup}, mass and internal energy are continuously injected into the wind launching region, as indicated by the lightly shaded area around the WD surface. An aligned rotating magnetic dipole is situated within the WD, and the resulting magnetic field lines are represented by the solid lines. 
Since the plasma beta (top-right panel of Fig.~\ref{fig:wind_profile}) at the WD surface is significantly smaller than unity, the injected gases co-rotate with the magnetic field up to approximately the Alfv$\acute{\text{e}}$n radius, $r_{\mathrm{A}}$, shown with the dotted line; 
we determine $r_{\mathrm{A}}$ from the condition $\rho(r_{\mathrm{A}}) |\mathbf{v}(r_{\mathrm{A}})|^2 = \left|\boldsymbol{B}(r_{\mathrm{A}})\right|^2 / (4 \pi r_{\mathrm{A}})$. 
As shown in the bottom-right panel of Fig.~\ref{fig:wind_profile}, the poloidal component of the magnetic field dominates inside the  Alfv$\acute{\text{e}}$n radius, maintaining the dipolar structure. 
In this region, the gases acquire azimuthal velocities due to the magnetic centrifugal force. On the other hand, the gases are also accelerated by the thermal pressure gradient at the outer edge of the wind launching region, causing them to expand radially. 
As the magnetic field strength decreases more rapidly with radius than the inertia of the expanding gases, the magnetic field structure undergoes modification, and the toroidal component dominates outside the Alfv$\acute{\text{e}}$n radius.

In the quasi-steady state, magnetic fields are fully open in directions away from the equatorial plane ($\theta \lesssim 80^{\circ}$ and $\theta \gtrsim 100^{\circ}$), where the wind is primarily accelerated by the pressure gradient at the outer edge of the wind launching region and becomes supersonic at $r \sim 2 \,R_*$.
In Fig. \ref{fig:wind_profile}, the sonic radius $r_\mathrm{s}$ is depicted with the dashed line; 
where we determine $r_\mathrm{s}$ based on the condition $|\mathbf{v}(r_\mathrm{s})| = c_\mathrm{s}(r_\mathrm{s})$. 
Note that the terminal velocity is comparable to the escape velocity (as shown in the bottom-right panel of Fig.~\ref{fig:wind_profile}), and the azimuthal velocities are at most a few percent of the radial velocities. 
Therefore, the properties of the wind in these directions are broadly consistent with the non-magnetized spherical Parker wind, 
even though the plasma beta at small radii is significantly less than unity.

In the equatorial direction ($80^{\circ} \lesssim \theta \lesssim 100^{\circ}$), magnetic fields are closed at small radii, forming a corotating magnetosphere. 
Beyond the last closed loop, the magnetic field lines are open with a predominant toroidal component, having opposite polarities with respect to the equatorial plane.
The transition of the magnetic field configuration is mediated by reconnection occurring at around the tip of the last closed loop, or the Y point.
As can be observed from the top-right panel of Fig.~\ref{fig:wind_profile}, the plasma beta in this transition region is higher than those along the open magnetic fields, implying that the gas is trapped mainly by magnetic tensions. In this high plasma-beta region sandwiched by low plasma beta regions, gases are pinched and radially accelerated in the reconnection region, eventually become supersonic at around $r \sim 5 \,R_*$.

We confirm that the terminal velocity of the fastest portion becomes comparable to the Michel velocity, $v_{\mathrm{M, \max}} \sim \sigma^{1/3} v_\mathrm{esc}$. 
The azimuthal velocity at the Y point is 30\% of $v_\mathrm{esc}$, which roughly corresponds to the corotation velocity at that location. After becoming ballistic, it gradually decreases as $\propto 1/r$ due to angular momentum conservation. 

\begin{figure}
        \centering
        \includegraphics[width=0.8\textwidth]{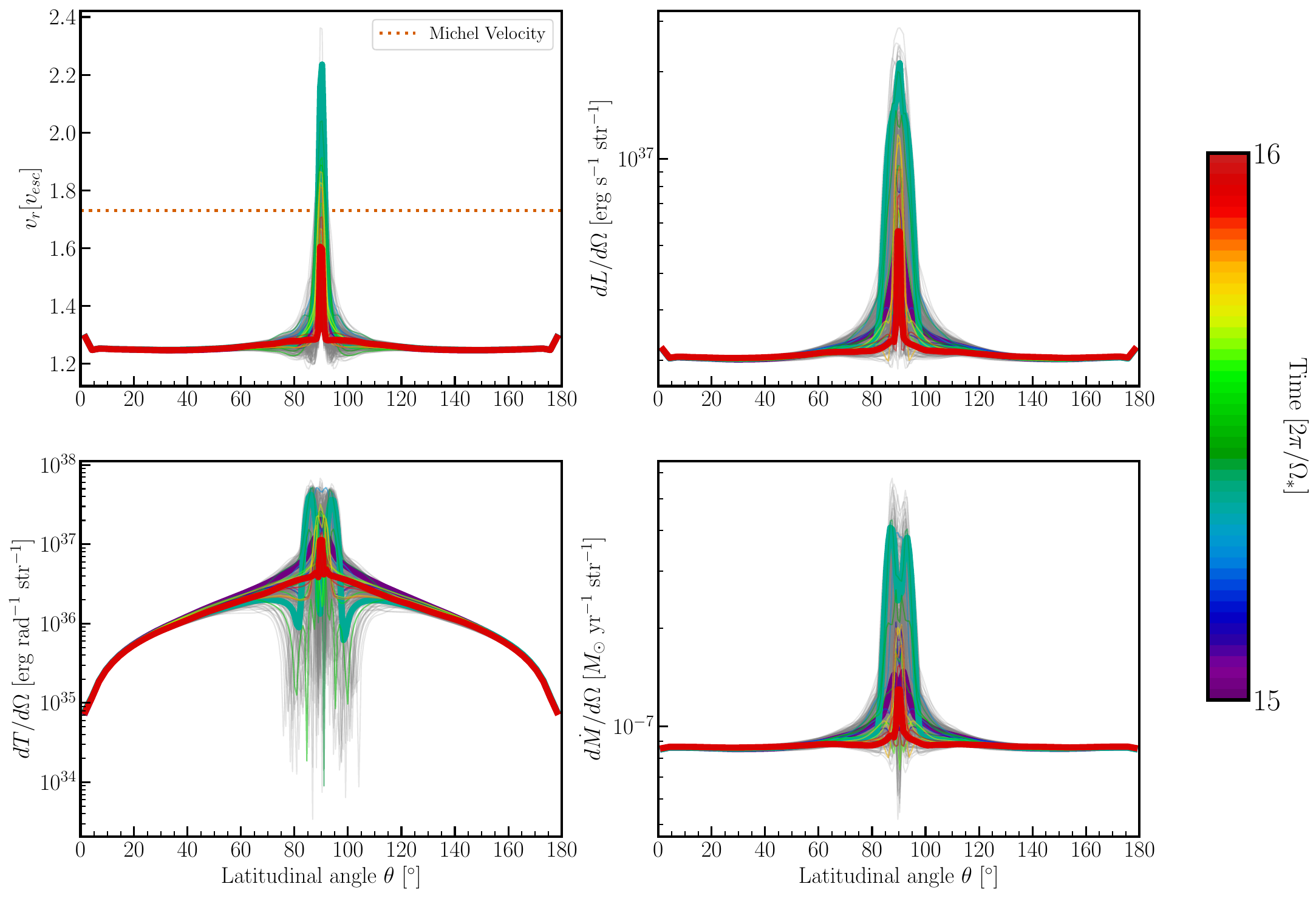}
        \caption{
        Latitudinal angle dependence of the rotating magnetic wind of $\tt B1.5e6\Omega0.23$ showing the radial velocity $v_r$ (top left panel), the wind luminosity $L$ (top right panel), the torque $T$ (bottom left panel), and the mass loss rate $\dot{M}$ (bottom right panel) at the outer boundary.
        A time sequence during a rotation period ($t=15\mbox{-}16$ [$2 \pi / \Omega_*$]) is represented with colors across the gray shaded region, indicating the entire dynamic range during the simulation.
        The thick purple, green, and red lines highlight the timings of pre-eruption, eruption, and post-eruption, respectively. 
        These timings are marked with vertical dashed lines in the upper panel of Fig. \ref{fig:time_evo}. 
        The orange dotted line in the top left panel indicates the Michel velocity (Eq. \ref{eq:WD}).
        }
        \label{fig:angle_dep}
    \end{figure}

The latitudinal angle dependence of the wind at the outer boundary is illustrated in Fig. \ref{fig:angle_dep}, where the gray shaded regions represent the dynamical range of radial velocity (top-left), luminosity (top-right), torque (bottom-left), and mass loss rate (bottom-right).
As described in the previous paragraphs, radial velocities in the near-equatorial direction can reach and transiently exceed the Michel velocity (represented by the horizontal dotted line) during eruptions caused by reconnection at the Y-point. Consequently, the wind luminosity, dominated by the radial kinetic term, also exhibits a sharp peak in the equatorial direction. On the other hand, the torque is primarily exerted by corotation with the magnetic fields and reaches its peak slightly off the equatorial plane ($\theta \sim 85^{\circ}$), corresponding to the edge of the concave shape of the Alfven radius.

\subsection{Time variability} \label{EML}

\begin{figure}
        \centering
        \includegraphics[width=1.0\textwidth]{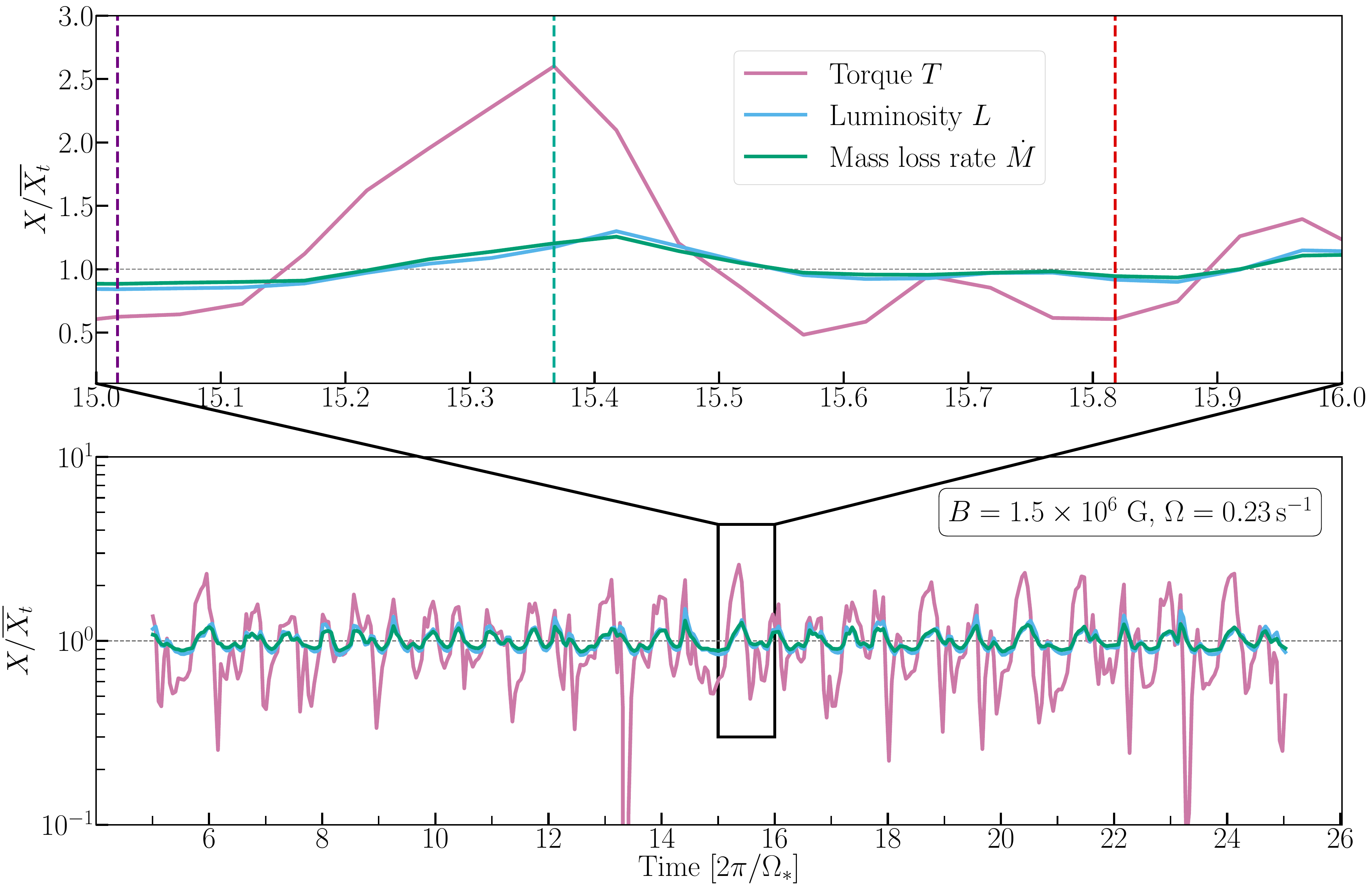}
        \caption{Time evolution of mass loss rate $\dot{M}$, luminosity $L$, and torque $T$ of the rotating magnetic wind of $\tt B1.5e6\Omega0.23$ estimated at the outer boundary of the computational domain. 
        The quantities are normalized by the time-averaged values. 
        The upper panel displays a close-up view of a rotation period ($t=15\mbox{-}16$ [$2 \pi / \Omega_*$]), where the vertical dashed lines indicate the timings of pre-eruption, eruption, and post-eruption highlighted in Fig. \ref{fig:angle_dep}.
        }
        \label{fig:time_evo}
\end{figure}

The acceleration of the rotating magnetic wind in the equatorial direction occurs in a time-variable manner, associated with magnetic reconnection at the Y-point. 
Consequently, the overall flux of mass, energy, and angular momentum from the central WD can also vary with time.
Fig. \ref{fig:time_evo} displays the time evolution of the mass loss rate ($\dot M$), luminosity ($L$), and torque ($T$) in our fiducial model. All quantities are normalized by their time-averaged values.

The lower panel of Fig. \ref{fig:time_evo}, which provides a long-term perspective, reveals a recurrent eruptive behavior.
A recurrent cycle consists of pre-eruption, reconnection, post-eruption phases: 
In the pre-eruption phase, gases injected into the near-equatorial plane become trapped within the closed field lines. 
Due to the centrifugal force, the gases accumulate at the tip of the last closed loop, resulting in a continuous decrease in plasma beta in that region. 
When the centrifugal force acting on the accumulated gases exceeds the tension of the closed magnetic fields, the tip of the closed zone starts expanding radially and is subsequently ejected as a plasmoid through reconnection.
Such a plasmoid can be observed at $r \sim 4 \, \mbox{-} \, 5 \, R_*$ in Fig. \ref{fig:wind_profile}.
Afterwards, the cycle returns to the pre-eruption phase and restores the gases within the closed magnetic field lines.
This type of recurrent eruptions has been known as slingshot prominence in the context of magnetically-active rapidly-rotating stars~\citep[e.g.,][]{2000MNRAS.316..647F,2005MNRAS.357..251T,2019MNRAS.482.2853J}. 

The upper panel of Fig. \ref{fig:time_evo} focuses on the fluxes during a rotation period ($t=15\mbox{-}16$ [$2 \pi / \Omega_*$]), as also depicted in Fig. \ref{fig:angle_dep}. 
In comparison to the pre- and post-eruption phases, represented by the purple and red lines, respectively, the observed fluxes of mass, energy, and angular momentum consistently increase as the erupted plasmoids reach the outer boundary, indicated by the green lines. 
Notably, the magnetic torque significantly contributes to the overall torque increase during the eruption phase and plays a dominant role in the central WD's spin-down.

We note that the specifics of reconnection dynamics, such as the frequency of recurrent eruptions and the resulting time evolution of mass, energy, and angular momentum fluxes, may be influenced by our numerical parameters, including spatial resolution (which governs numerical resistivity) and the width of the wind launching region. However, we have verified that the time-averaged values of wind velocities, mass loss rate, luminosity, and torque have all reached convergence concerning the spatial resolution in our simulations and the width of the wind launching region (see Appendix \ref{appen:convergence}).

\subsection{Scaling relation of the spin-down torque}
\label{spindowntorque}

Here we consider how the spin-down torque of rotating magnetic wind depends on the system parameters based on our numerical results. 
As we show in the previous sub-sections, the time-averaged torque is essentially determined by the magnetic torque exerted on the gases at the tip of the last closed field lines, or the Y-point, where the field configuration is still roughly compatible with the rotating dipole. 
In this case, the (electro)magnetic torque at the Y point can be estimated as
\begin{equation}
    \label{general_torque}
    T \approx \frac{\mu^2_*}{r^3_\mathrm{Y}},
\end{equation}
where $r_\mathrm{Y}$ represents the Y-point radius. 
Eq. (\ref{general_torque}) is based on the analogy with the force-free limit~\citep[e.g., ][]{2006ApJ...643.1139C}. 
In the force-free limit, the last closed field line corresponds to the light cylinder, $r_\mathrm{Y} \approx r_\mathrm{lc} = c/\Omega_*$, and the spin-down torque of a rotating dipole is roughly given as $T_\mathrm{ff} \approx B_\mathrm{lc}^2 r_\mathrm{lc} \approx [B_* \times (r_\mathrm{lc}/R_*)^{-3}]^2 R_\mathrm{lc}^3 \approx \mu_*^2/r_\mathrm{lc}^3 \approx \mu^2_*/r^3_\mathrm{Y}$. 

Based on our numerical results, the Y-point radius is determined by the balance between the centrifugal force and the magnetic tension force excerted on the gases, which can be described as 
\begin{equation} \label{eq:force_eq}
\rho_\mathrm{Y} r_{\rm Y} \Omega^2_*  \approx B^2_{\rm Y} \kappa_\mathrm{Y},
\end{equation}
where $\rho_{\rm Y}$ is the density, $B_{\rm Y}$ is the magnetic field streangth, and $\kappa_\mathrm{Y}$ is the curvature of the magnetic field at the Y point. 
In our case, the mass injection from the wind launching region is designed to be spherical, allowing us to describe the density at the Y point as 
\begin{equation}\label{eq:rho_Y}
    \rho_{\rm Y} \approx \frac{\dot{M}}{4 \pi  v_{r,\mathrm{Y}} r^2_{\rm Y}} \approx \frac{\dot{M}}{4 \pi  v_\mathrm{esc} r^2_{\rm Y}}.
\end{equation}
For the latter equation, we take into account that the gases at the Y point is quasi-hydrostatic in the radial direction and $v_{r,\mathrm{Y}} \approx v_\mathrm{esc}$ is satisfied in all cases examined in Table~\ref{table:3}.
On the other hand, 
given again that the magnetic field configuration at around the Y point is still roughly compatible with the rotating dipole, the strength and curvature of the magnetic field can be estimated as 
\begin{equation}\label{eq:B_Y}
    B_\mathrm{Y} \approx B_* \left(\frac{R_*}{r_\mathrm{Y}}\right)^3,
\end{equation}
\begin{equation}\label{eq:k_Y}
    \kappa_\mathrm{Y} \approx \frac{r_{\rm Y}}{R^2_*},
\end{equation}
respectively.
By substituting Eqs. (\ref{eq:rho_Y}–\ref{eq:k_Y}) into Eq. (\ref{eq:force_eq}), the Y-point radius can be obtained as
\begin{equation} \label{ry}
r_{\rm Y} \approx \left(\frac{\Phi_*^2 v_\mathrm{e s c}}{\Omega_*^2 \dot{M}}\right)^{1 / 4}.
\end{equation}
Using Eq. (\ref{eq:torque}) and the dimensionless parameter $\sigma$, the torque of a rotating magnetic wind can be expressed as 
\begin{equation}
\label{eq:scaling}
    \mathcal{T} \equiv \frac{T}{\dot{M} \Omega_* R^2_*} \approx \sigma^{1/4}. 
\end{equation}

\begin{figure}
    \centering
    \includegraphics[width=0.6\textwidth]{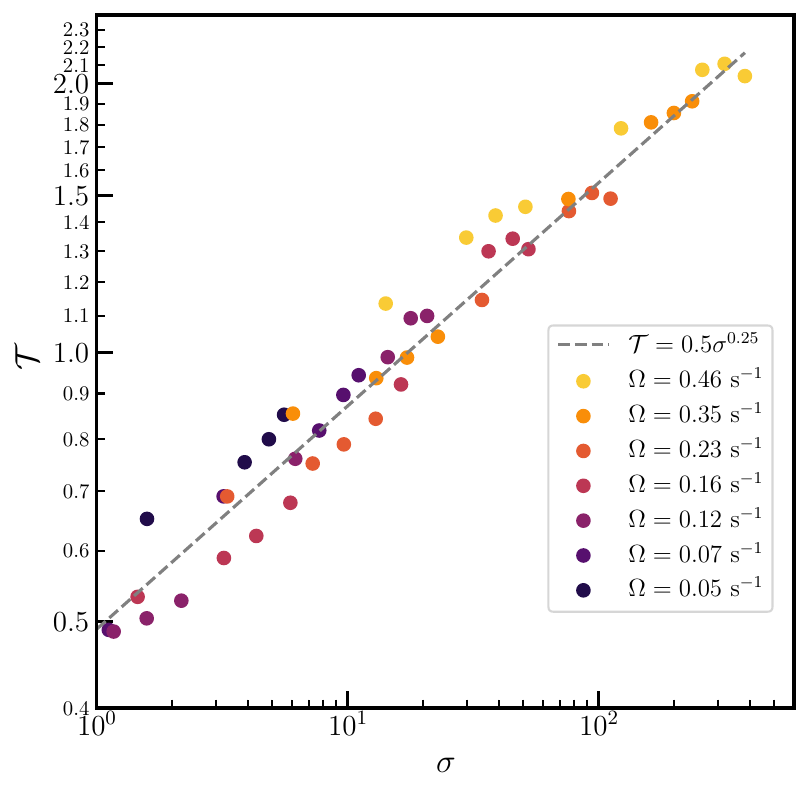}
    \caption{Scaling of the dimensionless time-averaged spin-down torque ($\mathcal{T} \equiv T/\dot M \Omega_* R_*^2$) with respect to the dimensionless parameter $\sigma \equiv \Phi^2_* \Omega_*^2/\dot M v_\mathrm{esc}^3$. 
    Each color of the points corresponds to an angular frequency ranging from $\Omega_* = [0.05,0.07,0.12,0.16,0.23,0.35,0.46]\,\mathrm{s^{-1}}$.
    The dashed line represents a fitting formula, $\mathcal{T} = 0.5\times \sigma^{1/4}$.}
    \label{fig:scaling}
\end{figure}

Fig. \ref{fig:scaling} shows the relation between the dimensionless parameters $\sigma$ and $\mathcal{T}$. 
While the above derivation of Eq. (\ref{eq:scaling}) is crudely approximate, the derived scaling relation is broadly consistent with our simulation results. The data points, regardless of their angular frequencies, can be effectively fitted by a single relation: $\mathcal{T} = 0.5 \times \sigma^{1/4}$.  
With restoring the physical dimensions, we obtain a fitting formula for the time-averaged spin-down torque of the rotating magnetic wind as 
\begin{equation} \label{eq:absoluteT}
    T \approx \frac{\dot{M} \Omega_* R^2_* \sigma^{1/4}}{2} \sim 2.4 \times 10^{36} \, \mathrm{erg}\,\mathrm{s}^{-1}  \left(\frac{M_*}{M_\odot}\right)^{-3/8} \left(\frac{R_*}{0.009 R_\odot}\right)^{27/8} \left(\frac{\dot{M}}{10^{-6} \, M_\odot \, \mathrm{yr}^{-1}}\right)^{3/4} \left(\frac{B_*}{10^6 \, \mathrm{G}}\right)^{1/2} \left(\frac{\Omega_*}{0.1 \, \mathrm{s}^{-1}}\right)^{3/2},
\end{equation}
which can be applicable at least to the cases with $1 \lesssim \sigma \lesssim 10^3$.

\section{Summary and Discussion} \label{sec:dis}

We have conducted a series of axisymmetric MHD simulations for rapidly rotating and strongly magnetized WDs, taking into account a near-surface carbon burning process as observationally inferred for WD J005311. 
We systematically investigated the wind anisotropy, time variability, and the spin-down evolution with respect to the dimensionless parameter $\sigma$ (Eq. \ref{sigma}).
We have confirmed that a co-rotating magnetosphere forms beyond the wind launching region and inside the Alfv$\acute{\text{e}}$n radius for $\sigma \gtrsim 1$, which leads to an anisotropic wind structure. In the near-equatorial directions there happens recurrent eruptions of plasmoids that are triggered by reconnections near the Y point. 
These plasmoids are accelerated to a radial velocity compatible with the Michel velocity, 
while the wind properties remain broadly consistent with the Parker wind away from the equatorial plane.
We found a scaling relation for the spin-down torque (Eq. \ref{eq:scaling}) that can be consistently explained by the criteria for reconnections to happen around the Y point, based on the numerical facts we obtained.
Our results complement previous studies on solar-like stars with relatively slow rotation~\citep[e.g., ][]{2009MNRAS.392.1022U,2012ApJ...754L..26M,2023arXiv230205462R}, and can be applied to not only massive WD merger remnants, but also various stellar objects with $1 \lesssim \sigma \lesssim 10^3$.

We now discuss implications of our numerical results on the properties of WD J005311. 
To reproduce the observed maximum wind velocity of $v_{\infty}=16,000 \pm 1,000 \mathrm{~km} \mathrm{~s}^{-1}$ by the rotating magnetic wind near the equatorial plane, and considering the carbon burning to be occurred near the WD surface, the WD paramters are constrained to be $M_* \sim 1.1\mbox{-}1.3\,M_\odot$, $B_* \sim (2\mbox{-}5)\times 10^7\,\rm G$, and $\Omega \sim 0.2\mbox{-}0.5 \,\rm s^{-1}$~\citep{2019ApJ...887...39K}.
Consequently, the dimensionless parameter ranges from $\sigma \sim 2\mbox{-}3$. 
\begin{itemize}
    
\item Given the Michel velocity to be $v_\mathrm{M, max} \approx \sigma^{1/3} v_\mathrm{esc}$, the contrast in radial velocity between the equatorial and polar directions is $\sim \sigma^{1/3} \sim 1.2\mbox{-}1.4$.
Such an anisotropic velocity profile could manifest in the optical spectrum.
To identify this signature, a multi-dimensional line transfer calculation based on our optically-thick rotating magnetic wind solution is necessary.

\item Assuming the mass injection from the carbon-burning region to be spherical, the time-averaged mass loss is also presumed to be spherical.
In other words, the quantity $\rho v_r$ remains relatively constant concerning the latitudinal angle. 
Consequently, the difference in wind ram pressure, which is proportional to $\rho v_r^2$, between the equatorial and polar directions is estimated to be $\sim \sigma^{1/3}$, roughly within the range of 1.2-1.4.
This can result in a non-spherical expansion of the wind termination shock.  
The wind nebula of WD J005311 has recently been shown to have an extended structure by {\it Chandra}~\citep{2023arXiv230414669K}.
Continued observations might identify any asymmetry or non-spherical characteristics.

\item The reconnection around the Y-point occurs in a time-dependent manner, which makes the wind acceleration and the resultant non-thermal radiation also time-variable. However, given that the Y-point is well within the photosphere in the case of WD J005311 (with $r_\mathrm{ph} \sim 0.15 R_\odot$) and the light crossing time at the wind termination shock is much longer than the expected reoccurrence time of reconnection, any time variability induced by reconnection may become smeared out and is difficult to detect. This can potentially explain the absence of apparent variabilities in WD J005311.

\item Using the scaling relation for the spin-down torque (Eq. \ref{eq:absoluteT}), we can estimate the spin-down timescale of WD J005311 as $t_\mathrm{sd} \approx 2 M_* R^2_* \Omega_*/5T$, or
 \begin{equation}
     t_\mathrm{sd} \sim 7.2\times 10^4 \, \text{yr}
    \left(\frac{M_*}{1.2\,M_\odot}\right)^{11/8} \left(\frac{R_*}{4,000\,\mathrm{km}}\right)^{-11/8}\left( \frac{\dot{M}}{3\times 10^{-6} \, M_\odot \, \mathrm{yr}^{-1}}\right)^{-3/4} \left(\frac{B_*}{2\times 10^7 \, G}\right)^{-1/2} \left(\frac{\Omega_*}{0.2 \, \mathrm{s}^{-1}}\right)^{-1/2}. 
 \end{equation}
Hence, even if the currently observed wind of WD J005311 is a rotating magnetic one and continues to blow for a Kelvin-Helmholtz timescale of the central WD, which is $\sim 1,000\mbox{-}10,000,\mathrm{yr}$, the spin-down will be negligible.
When the carbon burning in the near-surface region ceases, the mass loss rate will significantly decrease, which increases the dimensionless parameter $\sigma$. The rotating magnetic wind will then become relativistic and eventually enter the force-free regime without significantly spinning down the WD. In this case, the remnant WD may serve as a non-thermal radiation source, or or the so-called WD pulsar~\citep[e.g.,][]{2011PhRvD..83b3002K}.

\end{itemize}

Finally, we address some caveats in our numerical simulations. 
We have implemented a simple prescription for the near-surface carbon burning region as source terms (Eqs. \ref{eq:source_term_rho} and \ref{eq:source_term_e}), referred to as the wind launching region. 
However, the actual near-surface carbon burning region should be convective, and can be affected by the strong magnetic field. 
The structure of the convective region, the resulting wind launch, and its chemical composition would also be influenced by the radiative transfer.
For accurate multi-wavelength spectrum calculations, it is desirable to conduct a comprehensive radiative MHD simulation that covers from the carbon burning layer to the photosphere radius. 
Also, we only investigate the aligned rotating dipole magnetic fields in this paper, while a more complicated field configuration such as oblique or off-centered dipole may be realized for the remnant WD system.
Finally, the deformation of the central WD due to its rapid rotation and anisotropic carbon burning can alter the observed properties as well. We save the investigations into the above topics for our future work.


\begin{acknowledgments}
YZ is supported by the International Graduate Program for Excellence in Earth-Space Science (IGPEES) at the University of Tokyo. 
This work is also supported by Grants-in-Aid for Scientific Research No. JP23KJ0392(YZ), JP20K04010, JP20H01904, JP22H00130(KK), JP21H04487, JP22KK0043, JP22K14074 (ST), JP22K03688, JP22K03671, JP20H05639 (TS), and JP20K14512(KF). 
We thank Eliot Quataert for fruitful discussions and useful suggestions.
\end{acknowledgments}

%


\software{Athena++ 
          }



\appendix 

\section{Dual energy formalism} \label{appen:dual_e}

We introduced the so-called dual energy formalism to treat the magnetically dominated region in our simulations.
This method was originally developed by \cite{1995CoPhC..89..149B}, in order to deal with simulations with high Mach number flow. The basic idea is to separately solve the equation of internal energy in high Mach number ($\mathcal{M}$) region, and smoothly connect it to the solution given by the equation of total energy while $\mathcal{M} \sim 1$. We applied it to our case, where the magnetic energy (instead of the kinetic energy in \cite{1995CoPhC..89..149B}) dominates over others especially around the WD surface, so the relevant parameter is now plasma $\beta$ instead of $\mathcal{M}$. Details are given as follows.

First of all, internal energy $e_\mathrm{in}$ can be written in terms of the kinetic energy $\text{E}_k$, the total energy $\text{E}_\mathrm{tot}$ and magnetic energy $e_m$ as
\begin{equation} \label{ein_cons}
    e_\mathrm{in} = \text{E}_\mathrm{tot} - e_\mathrm{m} - \text{E}_\mathrm{k}
\end{equation}
in our simulations. Considering the case with magnetic energy dominated ($\beta \lesssim 1$), right-hand side of this equation becomes a difference between two large numbers, which is problematic for numerical computations, and becomes worse with $\beta$ decreases.

By solving the internal energy equation
\begin{equation} \label{ein_evo}
\frac{\partial e_\mathrm{in}}{\partial t}+ \mathbf{v} \cdot \nabla e_\mathrm{in}= -p \nabla \cdot \mathbf{v}
\end{equation}
separately, we can take the solution as a floor to prevent randomly small or even negative numbers that can possibly appear from Eq.(\ref{ein_cons}) in magnetically dominated region, and a smooth transition should be made at $\beta \sim 1$ to restore the solution given by Eq.(\ref{ein_cons}) while $\beta \gtrsim 1$. To achieve this, we define an effective internal energy in our simulations
\begin{equation}
    e_\mathrm{in, eff} \equiv \max\left[E_\mathrm{tot}-\frac{\rho|\mathbf{v}|^{2}}{2} - \frac{B^2}{2}, \, \eta_2\left(\frac{e_\mathrm{in}}{e_m}\right) e_\mathrm{in}\right],
\end{equation}
which depends on the ratio between internal energy given by Eq.(\ref{ein_evo}) and the ratio between internal energy and magnetic energy $e_\mathrm{in}/e_\mathrm{m}$. Here we choose the function $\eta_2(x)$ as
\begin{equation}
\eta_2(x) = \begin{cases} 0.99, & \frac{x}{x+0.03} < 0.99 \\ \frac{x}{x+0.03}, & 0.99 \leq \frac{x}{x+0.03} < 1 \\ 1, & 1 \leq \frac{x}{x+0.03} \end{cases},
\end{equation}
following \cite{2019ApJ...878L..10T} and \cite{2016PhDT.........5I}. This gives a safe enough internal energy floor as $0.99 \times e_\mathrm{in}$ at low plasma beta region. We then calculate the pressure from this effective internal energy before integrating the source term.

\section{Convergence of results} \label{appen:convergence}

\begin{figure}
        \centering
        \includegraphics[width=0.8\textwidth]{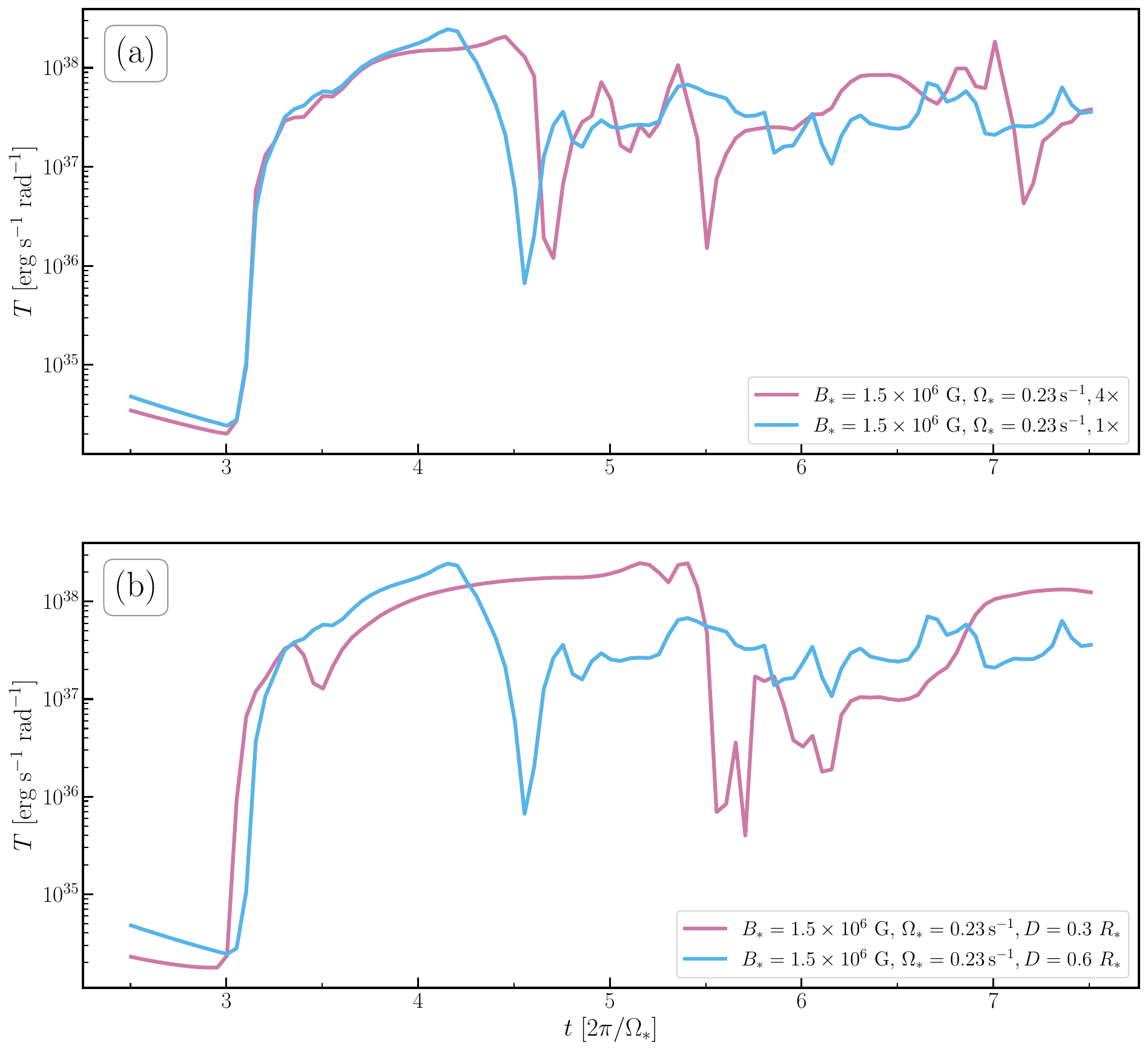}
        \caption{Time evolution of spin-down torque before and after (a) increasing resolution (b) decreasing the size of wind launching region in our fiducial case ($\tt B1.5e6\Omega0.23$).}
        \label{fig:reso1}
    \end{figure}

The convergence of our rotating magnetic wind solutions against both mesh resolution and the size of wind launching region have been confirmed. We show the results for our fiducial model ($\tt B1.5e6\Omega0.23$) in Fig. \ref{fig:reso1}. In the top panel we increase the spatial resolution for 4 times both along the radial and latitudinal direction, while in the bottom panel we change the thickness of wind launching region from $\mathcal{D}=0.6$ $R_*$ (fiducial value we are using, corresponding to 9 cells) to  $\mathcal{D}=0.3$ $R_*$ (which corresponds to 5 cells). We check the time evolution of the spindown torque $T$ for both changes, and zoom into the first few eruptive peaks to show the difference clearly. We find that the time-averaged value as well as the power-law trend converge with respect to both the spatial resolution and the size of the wind launching region, but the time variability vary. This is due to the fact that the reconnections in our simulations are mainly modulated by numerical resistivities. 

\section{Change of the mass loss rate in MHD regime} \label{appen:mdot}

\begin{figure}
        \centering
        \includegraphics[width=0.6\textwidth]{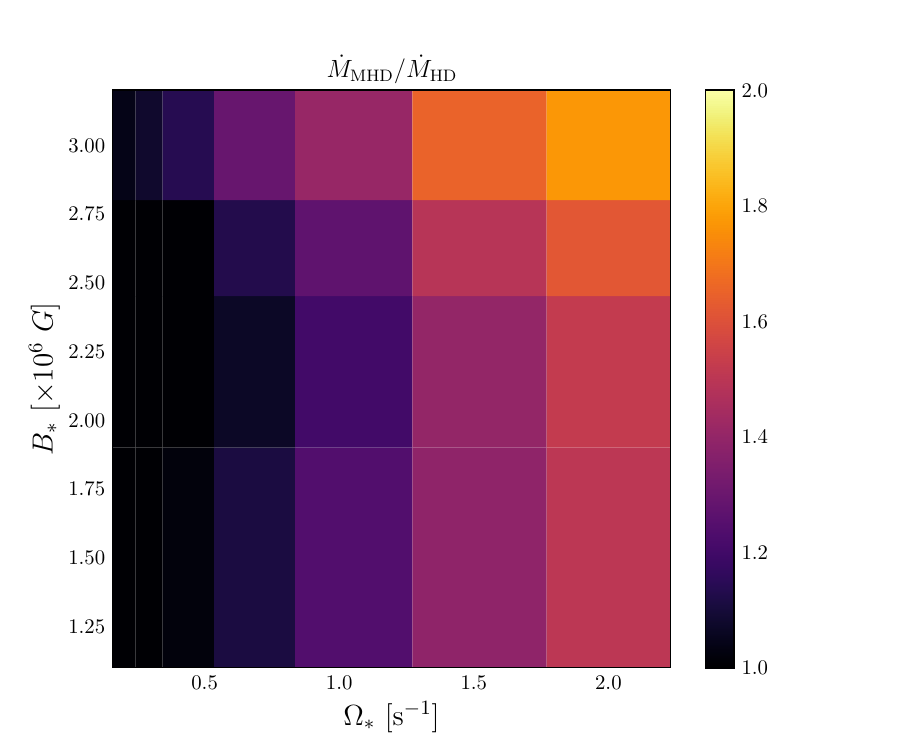}
        \caption{Parameter dependence of the ratio between time-averaged mass loss rate after turning on the magnetic fields $\dot{M}_\mathrm{MHD}$ and the value for pressure-driven wind $\dot{M}_\mathrm{HD}$ on surface magnetic field $B_*$ and WD spin frequency $\Omega_*$.}
        \label{fig:Mdot_B_omega}
    \end{figure}

As we claimed in Sec. \ref{sec:setup}, the mass loss rate is controlled to be the same for the pressure driven wind, based on the prescription of our wind launching region. However, for the rotating magnetic wind solutions we obtained, we found that time-averaged mass loss rate in MHD regime is slightly altered by increasing $B_*$ and $\Omega_*$, as shown in Fig. \ref{fig:Mdot_B_omega}. For all the cases we explored ($\sigma \sim 10^{0-3}$), $\dot{M}_\mathrm{MHD}/\dot{M}_\mathrm{HD}$ varies $\sim 2$ times at most, which is generally caused by the recurring eruption events (see the peak in the top right panel of Fig. \ref{fig:angle_dep}). Though minor in the regime we explored, it may get significant when beta further decreases, and the magnetic effects become increasingly important. 

\bibliography{ref}{}
\bibliographystyle{aasjournal}



\end{document}